\documentclass[pra,aps]{revtex4}
\usepackage{color}
\usepackage{ulem}

\newcommand{\bc}{\begin{center}}
\newcommand{\ec}{\end{center}\hrule}

\begin{document}
\title{G\"odel, Tarski, Turing and the conundrum of free will} %\date{}
\author{Chetan    S.    Mandayam   Nayakar}    \author{R.    Srikanth}
\affiliation{Poornaprajna    Institute    of   Scientific    Research,
  Sadashivnagar,  Bangalore}  \affiliation{Raman  Research  Institute,
  Sadashivnagar, Bangalore}

\begin{abstract}
The  problem of defining  and locating  free will  (FW) in  physics is
studied.   On basis  of  logical paradoxes,  we  argue that  FW has  a
meta-theoretic  character,  like  the  concept of  truth  in  Tarski's
undefinability theorem.  Free will exists relative to a base theory if
there is freedom to  deviate from the deterministic or indeterministic
dynamics  in the  theory,  with the  deviations  caused by  parameters
(representing will) in the  meta-theory.  By contrast, determinism and
indeterminism  do not require  meta-theoretic considerations  in their
formalization,  making FW  a fundamentally  new causal  primitive.  FW
exists relative to the meta-theory  if there is freedom for deviation,
due to higher-order causes.   Absolute free will, which corresponds to
our  intuitive  introspective notion  of  free  will,  exists if  this
meta-theoretic hierarchy  is infinite.   We argue that  this hierarchy
corresponds to  higher levels of  uncomputability. In other  words, at
any  finitely high  order  in the  hierarchy,  there are  uncomputable
deviations  from  the  law  at  that  order.   Applied  to  the  human
condition,  the   hierarchy  corresponds  to  deeper   levels  of  the
subconscious or unconscious mind.  Possible ramifications of our model
for physics, neuroscience and artificial intelligence (AI) are briefly
considered.
\end{abstract}
\maketitle

\section{Introduction}

Informally speaking,  FW is the  power to choose one  alternative from
many.  We  think that we have  it.  Yet its existence  and nature have
been debated  for over two thousand years  by philosophers, scientists
and theologians \cite{linda,tim}.  Free  will is, arguably, a familiar
stranger.   One basic difficulty  here is  its paradoxical  nature: FW
incorporates two  opposing notions: freedom  and control.  On  the one
hand,  freedom suggests indeterminism;  while control  suggests intent
and determinism.  From this perspective, FW is an oxymoron.  There are
some easy questions to ask about it that are hard to answer:
\begin{enumerate}
\item  \textit{What   is  FW?}   Physicists  typically  treat   it  as
  unpredictability or  uncorrelatedness with some past  data.  By this
  criterion,  uncontrolled or unreasonable  behavior would  qualify as
  free, and  thus does not  work.  On the  other hand, if  control and
  predictability are  the requirements, then  the altruism of  a saint
  and the  selfish actions of  a materialist are  equally predictable,
  though intuitively, the  former is the one we  would deem free.  The
  relevant question here is: how to define FW?
\item \textit{Does it exist is the world?} Introspection suggests that
  we are free in the sense that  we could have done other than what we
  did. But  unless FW  is well defined, this  intuitive feeling
  may well be illusory!
\item \textit{Is it  compatible with the laws of  physics?} The answer
  to this would depend on those to the above two questions!
\item \textit{What  is its neurological  basis, if FW exists?}
  Here again, the answer depends on the answer to the above three.
\end{enumerate}
This article  addresses the first  three questions above. The  last is
dealt  with  by  us  elsewhere  \cite{han}.   There  are  three  broad
metaphysical positions on FW: \textit{Compatibilism}, which holds that
determinism is compatible with FW.  A person may choose freely and yet
an  omniscient being may  possess foreknowledge  of that  choice.  The
opposite view  is \textit{incompatibilism}, according to  which FW and
determinism are incompatible.  Two divergent incompatibilist views are
\textit{hard determinism},  which rejects FW in  favor of determinism,
and  \textit{libertarianism}, which  rejects determinism  in  favor of
FW. To the determinist, free will  is at best illusory. The variety of
FW that arguably poses the greatest challenge is the liberatarian one,
and it is this that we consider in this work.

The  remaining   article  is   structured  as  follows.    In  Section
\ref{sec:qm}, we  briefly review  the current status  of free  will in
quantum mechanics. In Section \ref{sec:wfwp}, we present the Weak Free
Will paradox, which is concerned with question of the compatibility of
FW with  the laws  of physics. A  recapitulation of  a model of  FW we
developed in  Ref. \cite{nias2012} in  response to the Weak  Free Will
paradox is  presented in Section \ref{sec:fun}.  The  Strong Free Will
paradox, the  argument that the  only causal primitives in  Nature are
indeterminism   and    indeterminism,   is   presented    in   Section
\ref{sec:sfwp}.  In response to the  latter paradox, FW is proposed in
Section \ref{sec:meta}  as an infinite causal  hierarchy of meta-laws,
where higher-order laws can  cause deviations in lower-order ones. The
orders of the hierarchy are then differentiated by different levels of
uncomputability in Section  \ref{sec:compu}.  A justification for this
association  is  presented  in  Section  \ref{sec:link}.   The  fuller
implications for neuroscience, physics and AI are discussed in Section
\ref{sec:magic}. Finally, we conclude in Section \ref{sec:conclu}.
 
% theological 

\section{Quantum mechanics and FW \label{sec:qm}}

Quantum  mechanics,  by  introducing  the new  paradigm  of  intrinsic
randomness provided fresh impetus to the FW debate.  At first it seems
that quantum indeterminism provides room to accomodate free will. Yet,
randomness  also means loss  of control  \cite{vedral}.  Two  kinds of
freedom  are sometimes  discerned:  that of  experimenters who  choose
measurement  settings, and  of particles  that decide  the measurement
outcome  \cite{zei}.  T'  Hooft \cite{hoo07}  proposes  identifying FW
with freedom to modify the initial state of a system.

Reconsideration  of randomness  in  light of  quantum nonlocality  has
provided  new  insights.   Gisin  \cite{gisMWI} has  argued  that  the
Many-Worlds   interpretation  \cite{mwi}   of  quantum   mechanics  is
incompatible with  quantum nonlocality and  the existence of  FW.  The
Conway-Kochen FW Theorem \cite{ck,ck1} proves that given the violation
of Bell-type inequality and relativistic time-ordering, the assumption
of freedom of choice of  settings by observers implies a similar `free
will'  on the  part of  particles,  i.e., independence  from all  past
information  (with  `past' defined  as  the  past  light cone  or  all
complement of the future light cone).  Particle outcomes are then acts
of   creation   or  becoming   rather   than  pre-determinate   values
\cite{gisFWT}.    The  claim   \cite{ck,ck1}  that   the   freedom  of
experimenters  and the  violation of  Bell-type inequalities  rule out
covariant stochastic dynamical explanations of quantum nonlocality was
disputed by others \cite{tumu1,tumu2}.   A reconciliation of these two
positions    can    be    brought    about    by    the    observation
\cite{gisFWT,suFlash}   that  whereas   `collapse'  of   the  nonlocal
wavefunction leading  to correlated  outcomes is not  covariant, still
the `cloud  of future  events' is.  Hall  \cite{hall} has  refined the
notion  of FW  discussed  in the  Free  Will theorem,  by obtaining  a
trade-off between the freedom of observers or measurement independence
and the freedom of particles or indeterminism in measurement outcomes.

Suarez  \cite{suControl}  has  argued  that assuming  freedom  of  the
experimenter  in  a  measurement  configuration  where  two  observers
perform a  Bell inequality test \cite{chsh}, such  that each considers
her/his  own   measurement  as  having  been   performed  first,  then
time-ordering of  physical causation must  be given up.   Outcomes are
then  controlled  by  an  immaterial agency  outside  space-time  that
determines the order of  outcomes without affecting the probabilities,
given  by the  Born rule.   By this  argument, FW  and  randomness can
coexist \cite{suGFWT}.

\section{The Weak FW paradox \label{sec:wfwp}}

Consider  world $W$ governed  by \textit{deterministic}  physical law,
$L$. All  objects in $W$  evolve according to  $L$. The pair  $(W, L)$
constitute theory  $\mathcal{T}$.  By determinism,  the physical state
$\psi_t$ at time $t$ of an  inanimate agent (e.g., a particle) and the
physical  law  determine the  state  $\psi_{t+1}$  at subsequent  time
$t+1$:  $\psi_{t+1} =  L(\psi_t)$.  Free  will seems  to entail  for a
conscious agent $X$ in $W$, that in general,
\begin{equation}
\psi_{t+1} \ne L\left(\psi_t\right),
\label{eq:update}
\end{equation}
meaning that $X$'s  behavior is not captured by  $L$.  We may consider
replacing $L$ by a probabilistic law $L^\prime$ such that
\begin{equation}
\Psi_{t+1} = L^\prime\left(\Psi_t\right),
\label{eq:update0}
\end{equation}
where $\Psi_{t+1}$  is a random  variable representing probabilitistic
evolution,  characterized   by  mean  $\mu$   and  standard  deviation
$\sigma$.    By  the   Law   of  Large   Numbers,   the  sample   mean
$\langle\Psi_{t+1}\rangle^{(n)}$ of  the random variable $\Psi_{t+1}$,
over $n$ trials satisfies
\begin{equation}
\lim_{n     \rightarrow     \infty}     \textrm{Pr}\left(\left|\langle
\Psi_{t+1}\rangle^{(n)} - \mu\right|>\epsilon\right)=0,
\end{equation}
In other words, it is  exponentially unlikely that the agent $X$ will,
over many  trials, choose  \textit{atypical} sequences. There  is thus
long-run determinism and on that scale, lack of freedom.

Thus  libertarian FW is  compatible with  neither a  deterministic nor
indeterministic physical law. This is  the Weak Free Will paradox.  We
conclude that, libertarian FW, if it  exists for an agent $X$, must be
a  supra-physical commodity that  causes \textit{deviations}  from the
laws  of the  world  physical  $W$ in  which  $X$ (physically)  lives.
Causality  is no  longer  closed under  physics  in a  world where  FW
exists.  Otherwise-- if free-willed  action were explainable by $L$--,
then there is no deviation by definition, and FW must be illusory.

The   \textit{quantum  homeostatis  hypothesis}   \cite{suControl}  is
another response  to the  Weak Free Will  paradox. According to  it, a
conscious agent  may exert  FW by controlling  the order  of outcomes,
without contradicting the  probability law Eq. (\ref{eq:update0}).  In
order  not to  deviate from  the  long term  distribution required  by
$L^\prime$,   periods  of   deliberate,  conscious   action   must  be
compensated by other periods of uncontrolled behavior (e.g., sleep).

\section{Free will as the power of mind over matter
\label{sec:fun}}

A proposal that incorporates  will-driven deviations from the physical
law, as required  in our resolution of the Weak  Free Will paradox, is
the  Freedom-Understanding-Nature  (FUN)  model  \cite{nias2012}.   An
agent is assumed to be  equipped with two possibly opposing faculties:
Nature  ($N$),  determined by  physical  law  $L$  via brain  (limbic)
dynamics,   genetic   proclivities   and   instinctual   drives,   and
Understanding  ($U$),  determined  by  higher  cortical  faculties  of
morality  and compassion.   Here $N$  is a  manifestation of  $L$ from
which will-driven deviations may occur.  For example, $N$ may manifest
as a psycho-somatic desire that  urges the agent to procure a physical
object,  while  $U$  may   advise  restraint.   Freedom  ($F$)  is  an
immaterial resource that empowers $X$  to deviate choice $C$ away from
the  option urged  on by  $N$ towards  the one  advised by  $U$.  This
deviation cannot be captured by a Hamiltonian dynamics or a mechanical
model,  since these  exist within  $L$. The  \textit{base-law}  of the
deviation corresponds  to the known  physical law (here $N$),  but the
parameters  of  deviation  (here   $\varphi$  and  possibly  $U$)  are
associated with  the \textit{meta-law}.  Together  the meta-law (which
encompasses  $L$) and  parameters  associated with  it consistute  the
\textit{meta-theory}.

Free  will is  the power  to  conform to  $U$, deviating  from $N$  if
necessary.  A  simple mathamatical model:  suppose agent $X$  is faced
with a two-valued choice ``0''  or ``1''.  $U$ and $N$ are represented
by  random  variables $\hat{U}$  and  $\hat{N}$  that  take values  in
dichotomic choice  space of $\Omega  \equiv \{0,1\}$.  Let  $U$ advise
option  ``0''  whereas $N$  urges  $X$  towards  option ``1''  with  a
`strength' $\nu$.  The eventual choice $C$ of the agent is represented
by  random  variable  $\hat{C}$,  which  is a  convex  combination  of
$\hat{U}$ and $\hat{N}$, with the weight assigned to $U$ determined by
the  freedom  parameter  $\varphi$  (where  $0 \le  \varphi  \le  1)$,
representing $F$. Thus:
\begin{equation}
\hat{C} \equiv 
(1-\varphi)
\left(\begin{array}{c} 1-\nu \\ \nu \end{array}\right)
+
\varphi 
\left(\begin{array}{c} 1 \\ 0\end{array}\right)
=
\left(\begin{array}{c} \varphi +
(1-\varphi)(1-\nu) \\ (1-\varphi)\nu \end{array}\right)
\label{eq:fun}
\end{equation}
In  Eq.  (\ref{eq:fun}),  the larger  is $\varphi$,  the more  can $C$
deviate  from   $N$  towards  $U$.   If  freedom   is  maximal,  i.e.,
$\varphi=1$, then the agent behaves deterministically (mathematically,
$\hat{C}$   is  pure),   demonstrating  that   FW  is   distinct  from
unpredictability in  this model.  We note that  $\hat{C}$ becomes more
random as $\varphi$ drops from 1.

Free will $\Phi$ is quantified  as a measure of closeness of $\hat{U}$
to $\hat{C}$  normalized by the distance of  $\hat{U}$ from $\hat{N}$.
\begin{equation}
\Phi = \left(\vec{C}\cdot\vec{U}\right)|\vec{U}-\vec{N}|,
\label{eq:FW}
\end{equation}
where $\vec{C}$  ($\vec{N}$) is  a vector representation  of $\hat{C}$
($\hat{N}$) and  $|\cdot|$ denotes  trace distance. By  definition, $0
\le F \le 1$. For Eq. (\ref{eq:fun}), we find
\begin{equation}
\Phi = \frac{\varphi}{2}(\varphi + (1-\varphi)(1-\nu))|1-2\varphi|.
\label{eq:FW0}
\end{equation}
We may thus  think of FW $\Phi$ as innate  freedom $F$ (represented by
$\varphi$)  \textit{expressed}  physically when  there  is a  conflict
between  $U$  and $N$.  The  greater  the  conflict, the  greater  the
potential for  this expression.  Here innateness means  that $F$  is a
part of the meta-theory (or higher-order theory), while its expression
is at the base, physical level. 

\section{The strong FW paradox \label{sec:sfwp}}

That libertarian FW produces deviations  from the base law, only tells
how FW  acts.  It  does not  clarify what this  commodity is.   We now
attempt to define FW as the resource that can produce such deviations.
The  Strong Free  Will  paradox,  given here,  is  concerned with  the
question of  whether these deviations themselves  are lawful according
to a  higher law.  Suppose  agent $X$ is  faced with a  situation that
presents a choice. By our resolution of the Weak Free Will paradox, in
free choice, physical  causation fails in $W$ in  that $X$ may deviate
from the physical law $L$. $X$ has FW relative to $W$.  The parameters
of this deviation  are not governed by $L$,  and therefore are objects
that  live in  the extended  world $W^+$.   One now  considers whether
these parameters  themselves are lawful or free-willed.  For agent $X$
to have genuine FW, these deviation parameters should have been freely
chosen.

One possibility  is that there  there are deterministic laws  $L^+$ in
$W^+$ that explain  the causes that produce the  deviation on the base
law $L$. We then have:
\begin{equation}
\psi_{t+1}^{+} = L^{+}\left(\psi^{+}_t\right),
\label{eq:update2}
\end{equation} 
where $\psi_j^{+}$  is the extended  version of $\psi_j$,  obtained by
adding elements  from $W^+$. For  example, suppose $W^+$  includes the
physical  objects  in $W$  as  well  as  other more  abstract,  mental
objects. Then  $\psi_t$ represents  the state of  agent $X$'s  body at
time $t$,  then $\psi_t^{+}$  is that of  his body-mind at  that time.
Here $\psi_{t+1}$  is implicitly determined  by $L^{+}$ acting  on the
enhanced state $\psi^+_t$. In this case,  there is no FW in $W^+$, and
in an absolute sense, no FW  at all, since the deviations from $L$ are
deterministically  determined  by   the  higher  law  $L^+$.   Another
possibility is  that there is  a probabilistic law  $L^{+\prime}$ that
governs the $W^+$ objects, but here again there is no FW, as argued in
Section  \ref{sec:wfwp}.  By  the  Weak Free  Will  paradox, which  is
adapted to $W^+$  agency, there is FW only if  there can be deviations
from law $L^+$.

The causes (or  parameters) of these deviations are  objects that live
in a yet  higher world $W^{++}$.  It is clear  that the above argument
\textit{relativizes},  i.e.,  one  can  recursively  apply  the  above
argument, shifting the  level from $W^+$ to $W^{++}$,  and beyond.  If
this recursion  is finite, then although  there is FW in  the sense of
the Weak Free Will paradox till that causal depth, still there is none
beyond.  What remains  just above that depth is  either determinism or
indeterminism. In  other words,  the Strong Free  Will paradox  is the
argument that  the only fundamental causal  primitives are determinism
and indeterminism. There  is no room for any  primitive like FW, which
is  seen  ultimately to  be  a  hierarchical  interplay of  these  two
primitives.

The quantum homeostatis hypothesis  \cite{suControl} and the FUN model
\cite{nias2012},  which provide  a resolution  to the  Weak  Free Will
paradox, do not  appear to get past the Strong  Free Will paradox.  In
the former case, the Strong Free Will argument applies to the resource
by which  an agent controls the  ordering of outcomes, and  in the FUN
model, it applies to the freedom paramter $\varphi$.

\section{Free will as infinite meta-theoretic construct \label{sec:meta}}

We propose  that the response  to the Strong  Free Will paradox  is to
simply  embrace  it!   We  suggest   that  it  is  in  the  nature  of
Consciousness that  it can support  the above causal recursion  to any
depth.   For convenience,  let the  base  (physical) laws  $L$ now  be
denoted by $L^{(0)}$ and the corresponding physical world $W$ in which
they  operate, by  $W^{(0)}$. The  usual laws  of physics  are order-0
laws, which govern the behavior of order-0 objects, which are physical
objects,  denoted  $\lambda^{(0)}$,  such  as the  physical  component
(body) of  sentient agents like humans.  The physical state  of such a
being is denoted $\psi^{(0)}$.

Volition- or  will-driven deviations (if they exist)  from the order-0
laws  are  caused, according  to  Eq.   (\ref{eq:update}), by  order-1
objects in  $L^+$, which we  denote now by $L^{(1)}$.   These objects,
denoted by $\lambda^{(1)}$, live in the larger world $W^+$, denoted by
$W^{(1)}$.   We  have $W^{(0)}  \subset  W^{(1)}$, $\lambda^{(0)}  \in
W^{(0)}$  and $\lambda^{(1)}  \in  W^{(1)} -  W^{(0)}$.  The  extended
state of  the free-willed being  $X$ is $\psi^{(1)} \in  W^{(1)}$ such
that $\psi^{(0)}  = \psi^{(1)} \cap  W^{(0)}$.  Similarly, will-driven
deviations   from  $L^{(1)}$   dynamics  are   attributed   to  causes
$\lambda^{++}  \equiv \lambda^{(2)} \in  W^{++} \equiv  W^{(2)}$.  The
law $L^{(1)}$  is a meta-law for $L^{(0)}$  in that it is  law that is
about and governs $L^{(0)}$.  The  objects in $W^{(1)}$ and the axioms
of $L^{(1)}$ thus constitute a meta-theory $\mathcal{T}^{(1)}$ for the
base theory $\mathcal{T}^{(0)} \equiv (W^{(0)},L^{(0)})$.

At each level $j$ (a  positive integer), the pair $(W^{(j)}, L^{(j)})$
constitute  the order-$j$ meta-theory  $\mathcal{T}^{(j)}$.  Formally,
this  system of  meta-theories $\mathcal{T}^{(j)}$,  can  be continued
indefinitely.  The meta-theoretic construction  of FW is the assertion
that  at  each  level $j$,  FW  at  that  level  is expressed  as  the
possibility  of  deviation from  $L^{(j)}$  caused by  order-$(j{+}1)$
objects  $\lambda^{(j{+}1)}  \in  W^{(j{+}1)}$.  \textit{Relative}  FW
exists at  level $K$ if there  are deviations from  all $L^{(j)}$ with
$j<K$.  If there  are no deviations in $L^{(K)}$, then  there is no FW
beyond  order-$K$. We  propose the  existence of  \textit{absolute} FW
exists in  the sense that relative  FW exists for all  finite $K$.  In
other words, no matter how deep we go tracking down causes, the causes
of  these  causes, the  causes  of causes  of  causes,  and so  forth,
although  our  explanatory  power  may increase,  still  there  remain
unexplained  deviations.  

We note  that determinism  and indeterminism lack  this meta-theoretic
character in that they can be described within a given level (say $L$)
of  the  dynamics.  Free  will  then is  a  new  causal primitive,  in
addition  to determinism and  indeterminism.  But  although it  is not
logically necessary  for indeterminism to be  meta-theoretic, still it
may be a function of higher-order causes. It differs from free will at
the same  order in  not being correlated  with the  immediately higher
order  meta-theory. Thus  indeterminism  is just  freedom without  the
will.  Quantum  fluctations are  then order-0 freedom,  while sentient
beings who deviate from $\hat{N}$  in the FUN model possess (at least)
order-0 free will.

If no  such correlated parameter  is evident in the  meta-theory, then
there  is  freedom alone,  and  to  ensure  that causality  holds,  we
conclude  that  the  causes   lie  in  meta-theoretic  depths  farther
down. This is a subtle but socially significant difference between the
free will  of sentient agents  and the freedom of  particles. Although
there  is  no  logical  necessity  to  do  so,  we  propose  that  all
indeterminism is freedom in the above sense.

A  problem  that merits  further  study is  the  question  of how  the
different meta-theories $\mathcal{T}^{(j)}$ may be demarcated. One may
consider that $\mathcal{T}^{(0)}$  corresponds to predicate logic, and
$\mathcal{T}^{(j)}$   ($j>0$)   correspond   to  higher-order   logics
\cite{hol}.   An example  of a  demarcation that  doesn't work  is the
proposal   that  $\mathcal{T}^{(0)}$   is   classical  mechanics   and
$\mathcal{T}^{(1)}$  quantum mechanics,  the idea  being  that quantum
mechanics causes deviations from  classical behavior.  We venture that
quantum  mechanics,  and  thus  also classical  mechanics,  belong  in
$\mathcal{T}^{(0)}$, the  base physical theory.   Intuitively, this is
because quantum mechanics can be  considered as a modification over an
indeterministic generalization  of classical mechanics.  Although this
modification   is   responsible   for   peculiarities   like   quantum
nonlocality, it  does not  seem to endow  quantum mechanics  with more
explanatory power, as we clarify below.

\section{Computability and causality \label{sec:compu}}

We suggest  that the  different levels of  theory $\mathcal{T}^{(0)}$,
the    meta-theory     $\mathcal{T}^{(1)}$,    the    meta-meta-theory
$\mathcal{T}^{(2)}$,  etc. are  distinguished in  different  levels of
\textit{computability} that they  support. Computability, which is the
capacity  to  effectively solve  a  problem,  is  a topic  studied  in
mathematical  logic  and computer  science.   Computability theory  is
concerned  with the  question  of whether  an  algorithm or  effective
procedure exists  to solve  a problem.  It  is known that  there exist
algorithmically uncomputable  problems, e.g., the  halting problem for
computer programs  (formally: Turing machines)  \cite{md1958}. One can
conceive  qualitatively  more  powerful  computers  (formally:  oracle
machines) that can  solve these problems, but must  contend with their
own  halting problem.   One can  continue  this exercise,  to build  a
hierarchy  of  ever  more  powerful oracle-machines  and  ever  harder
uncomputable    problems.     We    propose   that    the    different
$\mathcal{T}^{(j)}$'s  correspond  to  the  different  rungs  in  this
hierarchy.  

We can now make precise the sense in which FW is said to exist in this
model.  In computation theory, an \textit{oracle} for problem $A$ is a
hypothetical device that given any  instance of $A$ returns the answer
in finite  time.  If a Turing  machine with access to  this oracle can
solve every instance  of problem $B$, then $B$  is Turing-reducible to
$A$.  In other  words,  if $A$  is  computable, then  so  is $B$.   If
similarly, $A$ is  reducible to $B$, then $A$  and $B$ are equivalent.
A  Turing-degree  is  the  set  of  all  problems  that  are  mutually
Turing-equivalent.

The  set  of  all  computable  problems  is  a  Turing  degree,  which
quantifies the level of its algorithmic unsolvability. A higher Turing
degree  is  the  set  of  problems Turing-equivalent  to  the  halting
problem, for example, the problem of whether a statement can be proved
from the axioms  of set theory. This latter set  is not solvable (with
Turing  machines).   One  can   construct  higher  Turing  degrees  by
relativizing the proof  of the halting theorem for  Turing machines to
Turing machines with oracles.

Thus   the    different   meta-theories   correspond    to   different
\textit{Turing degrees},  with a  higher-order theory being  of higher
Turing    degree.     The    move    from    $\mathcal{T}^{(j)}$    to
$\mathcal{T}^{(j{+}1)}$  is  equivalent   to  a  \textit{Turing  jump}
\cite{tjump}, i.e., produces a unit shift in the Turing degree.

By this criterion, since the  quantum version of Turing machine is not
more  powerful   than  a   classical  machine  from   a  computability
perspective (and  even computational complexity  perspective), quantum
mechanics  belongs to  the same-order  theory as  classical mechanics,
namely $\mathcal{T}^{(0)}$. On the  other hand, a deterministic hidden
variable  theory  for quantum  mechanics  is  a  likely candidate  for
$\mathcal{T}^{(1)}$.   For  one,   it  can  deterministically  explain
deviations  from classical  mechanics  in terms  of hidden  variables.
Further, the  ability to manipulate this  subquantum information leads
to  super-Turing power  in a  computational complexity  sense (solving
NP-complete problems in polynomial time) \cite{val02}.

Our  assertion of  absolute FW  above can  now be  re-stated  thus: it
exists in  the sense that the  problem of determining  an agent's free
choice is of infinite Turing degree, i.e., for every oracle machine at
finitely high order, there will be uncomputable residual deviations in
a model of free choice.  In  other words, FW is not just uncomputable,
but infinitely so!

The  first   hint  that  the   this  linking  between   causality  and
computability  is   an  appropriate   direction  to  proceed   is  the
observation   that   the  notion   of   \textit{proof}  in   G\"odel's
incompleteness  theorem \cite{god} and  of \textit{truth}  in Tarski's
undefinability  theorem  \cite{tarski},  are,  in some  ways,  similar
meta-theoretic concepts.  G\"odel  incompleteness may be considered as
an  avatar of  Turing  uncomputability \cite{md1958},  as  we show  in
Section \ref{sec:link}.

\section{Truth, proof and freedom \label{sec:link}}

Imagine  a   human  agent  $X$  equipped  with   the  usual  cognitive
apparatuses of a  physical brain and something else  called mind.  The
physical aspect of  $X$ lives in world $W^{(0)}$  governed by physical
law $L^{(0)}$, while the higher aspects of $\psi^{(j)}$ are attributed
to the  mind entity.   Now $\mathcal{T}^{(0)}$ naturally  determines a
model of computation, and  correspondingly the limits of computational
complexity   and   computability,   in   the  world   $W^{(0)}$.    If
$\mathcal{T}^{(0)}$ is the physical  world, then the relevant model is
Turing machines (TMs) or their equivalent.  All TMs, including the TMs
representing agents  $X, Y, Z$ et  al. that live in  $W^{(0)}$, can be
enumerated  according to a  fixed scheme,  e.g., G\"odel  numbering or
Turing numbering. We  denote the corresponding numbers $\underline{X},
\underline{Y}, \underline{Z}$ and so on.

We now present a proof of G\"odel's theorem via the uncomputability of
a  problem that is  equivalent to  the halting  problem \cite{md1958}.
Suppose algorithm $\mathcal{A}$ exists that can predict in finite time
agent $X$'s binary choice upon input $i$:
\begin{equation}
\mathcal{A}(\underline{X}; \underline{i}) =
\left\{
\begin{array}{cc}
0 \Longleftrightarrow X(i) = 0 \\
1  \Longleftrightarrow X(i) = 1
\end{array}, \right.
\label{eq:2}
\end{equation}
i.e., the machine  outputs 0 (1) if $X$ outputs 0  (1) acting on input
$i$.  We construct the following program:
\begin{equation}
\mathcal{R}(p) = \left\{
\begin{array}{cc}
1 \Longleftrightarrow \mathcal{A}(p;p) = 0 \\
0  \Longleftrightarrow \mathcal{A}(p;p) = 1,
\end{array} \right.
\label{eq:3}
\end{equation}
where $p$ is a positive integer. By construction, $\mathcal{R}$ exists
if  $\mathcal{A}$ does.   Applying $\mathcal{A}$  to  $\mathcal{R}$ we
have from Eqs. (\ref{eq:2}) and (\ref{eq:3}):
\begin{equation}
\mathcal{A}(\underline{\mathcal{R}}; p) = \left\{
\begin{array}{cc}
1 \Longleftrightarrow \mathcal{A}(p;p) = 0 \\
0  \Longleftrightarrow \mathcal{A}(p;p) = 1.
\end{array} \right.
\label{eq:4}
\end{equation}
We obtain  a contradiction when we set  $p = \underline{\mathcal{R}}$,
which  is an  instance of  the diagonal  argument, first  pioneered by
Cantor.  The conclusion is  that a general, sound predictive algorithm
$\mathcal{A}$ is logically impossible  because it would help create an
algorithm that is  so powerful that, knowing its  own future, it could
act contradictory to its own prediction.

Therefore,  if  the  system  $F$  of  reasoning  embodied  by  TMs  in
consistent,      then       to      avoid      the      contradiction,
$\mathcal{A}(\underline{\mathcal{R}};  \underline{\mathcal{R}})$ loops
infinitely.    Therefore,   $\mathcal{R}(\underline{\mathcal{R}})$  is
undecidable within $F$-- a G\"odel sentence for the system. How do we,
as human  beings reading this,  know this truth of  an algorithmically
undecidable   statement  like  $\mathcal{R}(\underline{\mathcal{R}})$?
Can the human mind therefore be more powerful than TMs? According to a
much debated  anti-mechanist view due to  Lucas \cite{lukas,lucas} and
Penrose \cite{pen1,pen2}, it is.

An objection \cite{putnam} is based on G\"odel's second incompleteness
theorem.  The  reasoning that  led us  to the claim  that ``If  $F$ is
consistent,     then     $\mathcal{R}(\underline{\mathcal{R}})$     is
non-terminating'' can  itself be programmed  into the action of  a TM.
Thus  if  $F$  is in  fact  consistent,  then  the  proof of  its  own
consistency     must     be     uncomputable,    in     order     that
$\mathcal{R}(\underline{\mathcal{R}})$ be unprovable  in $F$.  In this
objection to the anti-mechanist view of the mind, even if human agents
were  consistent, they  would be  unable to  establish this,  and thus
attain to a meta-mathematical comprehension of what TMs can't. To this
view, the  following counter-objection may  be given: Suppose  a large
number $n$ of persons toss a  fair coin each, assuming that humans are
consistent  if  a  head  turns   up,  and  that  we  are  inconsistent
otherwise. Then it follows that if it so happens that human beings are
consistent,   the    understanding   that   $\mathcal{R}   (\underline
{\mathcal{R}})$ is  non-terminating would make  the mind will  be more
powerful than  TMs in the case of  those people in the  group for whom
($n/2$ in  number, with exponentially  high probability) a  head turns
up!  Clearly, we wouldn't want to power of the mind to depend the whim
of a coin!   (On the other hand, if humans  happen to be inconsistent,
then the above  argument fails.)  Thus it seems  that objections based
on   consistency  of   $F$   do  not   satisfactorily  undermine   the
anti-mechanist view.

Another  criticisim \cite{ben}  is  that  it may  not  be possible  to
humanly construct  a G\"odel sentence for something  so complicated as
the TM  representing a  human being. However  this objection  seems to
assume that the human may in principle be more powerful than a TM, but
not in practice.  The claim  for the disputing the anti-mechanist view
should,  it would  seem, not  depend on  limitations imposed  on human
beings like  mortality, emotionalism, fatigue, etc., unless  it can be
demonstrated that such a factor will necessarily (logically) intervene
to prevent a human from computing the G\"odel sentence in question.

Yet another  objection to the  anti-mechanist view is due  to Whiteley
\cite{hvit},  which is  that humans  would  themselves be  prone to  a
G\"odel sentence,  not unlike  TMs.  Consider the  following predicate
formula, formalized suitably:
\begin{center}
P($\xi$)  $\equiv$ ``\textit{$\xi$  cannot  consistently believe  this
  statement.}''
\end{center}
Now $P(X)$ is true or false. If true, then $X$ does not believe $P(X)$
when   presented   with  it,   making   him  \textit{incomplete}   for
disbelieving a true  statement.  If $P(X)$ is false,  then $\neg P(X)$
is  true; in  other  words, he  \textit{believes}  a false  statement,
making him  \textit{inconsistent}.  To avoid  inconsistency, we select
for  him  to be  incomplete.   This is  just  an  informal version  of
G\"odel's  theorem,  with  disbelief \cite{mackay60}  self-referenced,
rather than unprovability (Actually,  presenting $P(X)$ to $X$ creates
a  double  self-reference:  in  $X$  and in  the  resulting  statement
\cite{sri+}).   According  to  this  objection,  humans  have  similar
limitations  as those  attributed to  machines on  basis  of G\"odel's
theorem. 

$P(X)$ can be  consistently believed by agent $Y$,  and $P(Y)$ by $X$.
Now imagine  that $Y$  believes $P(X)$ and  informs this  statement to
$X$.  As predicted $X$ is  unable to believe $P(X)$, being consistent.
When $Y$ asks him if he  believes $P(X)$, $X$ replies in the negative.
When $Y$  asks him why, $X$ says  that he cannot say  exactly why, but
somehow  $P(X)$   feels  unconvincing.   Now  $Y$  happens   to  be  a
mathematician, and explains G\"odel's  theorem to $X$ enlightening $X$
as to why $X$ is unable to believe $P(X)$.  $X$ is now able to reflect
upon the  nature of his  understanding and the limitations  imposed by
the theorem.  He  trusts $Y$ enough to know  that $Y$ wouldn't believe
$P(X)$ unless it were true. He then has an `Ah ha' moment, and decides
that $P(X)$  is true after  all.  His brain still  disbelieves $P(X)$,
but somehow he  knows that $P(X)$ is true, though  maybe he can't help
acting as if it were false.  If $X$ could \textit{not} have this Ah-ha
moment, he  could well be a machine,  and in social terms,  he must be
some kind of  zombie or dull-witted person, not  a normal human! Thus,
the power of  intuitive grasp seems to be that $X$  can supply his own
meta-theory on the  fly, and it is this power  to meta-jump that seems
to come from nowhere but  the nature of consciousness, and divides man
and machine!

In other  words, $X$  is able to  \textit{meta-straddle}: he  can both
believe   and   disbelieve   $P(X)$  simultaneously,   without   being
inconsistent because  these states occur at two  different layers: the
disbelief at $W^{(j)}$ (possibly $j=0$) and the belief at $W^{(j+1)}$.
(Here we may recollect moments  where our eyes/brain are deluded by an
optical  illusion which  we know  is illusory;  or when  we  know that
something is  false at  a higher level,  and yet can't  help believing
it!)   This  conclusion is  consistent  with  our  demarcation of  the
meta-theories   $\mathcal{T}^{(j)}$   in   terms  of   uncomputability
occurring in the theory \cite{sri+}.

Now it is conceivable that the reasoning that led us to the conclusion
that if $X$ intuits the G\"odel sentence, and then meta-straddles, can
itself be  formalized or equivalently mechanized  somehow, for example
having a multi-tape TM with a denial on the lower tape, and acceptance
on  a higher  tape that  accesses an  oracle. So  this  order-1 oracle
machine  will  be our  new  model for  agent  $X$.   However, one  can
construct  a Whitely  sentence  $P^{(1)}(X)$ for  $X$  modelled as  an
oracle-1 machine,  and repeat the  above argument, for  $X$ perceiving
the truth of this  sentence. Relativizing this procedure, we construct
ever more powerful  oracle machines, going up the  jump hierarchy, and
iterating  farther  into transfinite  ordinals,  and even  uncountable
regular  cardinals \cite{tjump}. In  each case,  $X$ modeled  as these
machines still makes the intuitve jump, because if it didn't, we would
call  it a  (higher-order) zombie.   Whereas the  very character  of a
machine  must  be changed  as  we climb  up  the  jump hierarchy,  the
consciousness of  $X$ seems  essentially to be  the same thing,  as it
jumps up  or hops outside  endlessly.  It has the  magical infinitely,
meta-expansive plasticity  that seems to  be the essence  of conscious
understanding, as against mechanized assimilation by an automaton.

\section{Mind and magic: possible ramifications for AI, 
physics and neuroscience
\label{sec:magic}}

Without going  into details, we  may assume that the  relation between
the physical aspect and the mind  of human agents is causal both ways,
i.e., that  each influences  the other. We  may deduce that  there are
oracles located  in the  brain, that serve  as gateways  through which
$\lambda^{(1)}$  and  higher influences  can  flow  into the  physical
system, to produce deviations from $L^{(0)}$ laws, giving rise to free
will in the physical plane.   Since our conscious actions are arguably
not more  powerful than  TMs, it follows  that the influence  of these
brain  oracles must  occur  at a  subconscious  or unconscious  level.
Therefore,  the   only  failsafe  way  to   access  this  super-Turing
computational  capacity  involved in  a  person's  free  choice is  to
present  the  person   with  the  choice,  and  allow   him  to  react
spontaneously.   To use  a  Biblical  allusion, there  was  no way  to
compute what Adam's reaction would  be to the injuction forbidding him
to  eat  the apple  in  the  forest of  Eden,  except  by putting  him
physically  in the forest  and forbidding  him!  From  a computational
perspective,  every  instance of  free  choice  is  a miracle!   These
observations have various ramifications discussed below.

Our model  leads naturally to  a Cartesian mind-body  dichotomy, where
the   mind  is   indefinitely  layered   into  increasing   levels  of
uncomputability.  Since  the conscious  body/mind is no  more powerful
than  TMs,   the  entirety  of  the  human   Consciousness  cannot  be
consciously grasped through a  finitary reasoning process, and must be
done  intuitively. From  the perspective  of the  physical  world, the
actions  of a free-willed  agent are  not always  predictable, without
being   random.   Mathematically,   free  choice   can   be  described
probabilistically as  in Eq.  (\ref{eq:fun})  of the FUN  model, where
probability arises  through \textit{unknowability}.  Any  AI algorithm
or neural network system, being no more powerful than Turing machines,
can never  in general  simulate a free  choice of human  or presumably
even an animal or, in general,  any entity `vivified by a soul'.  Thus
Consciousness,  of which intuitive  thought and  FW are  aspects, must
also  be also  unsimulable,  in contradistinction  to  the premise  of
Strong AI.

It  seems  possible  that  phenomena  like savantism  or  the  pattern
recognition in the brain make use of the higher computational capacity
attributed  to the  subconscious mind  in  our model.   There must  be
something  special  in the  structure  of  the  brain that  allows  an
interface  with  a  subconscious  oracle,  a  point  that  we  discuss
elsewhere \cite{han}.  We venture that the above phenomena are windows
on that structure.

The free-willed human being is not an isolated entity, but part of the
universe.  If  FW has  the above implications  for the physics  of the
human  brain,  this  must   have  some  implications,  significant  or
otherwise,  for  the  physics  of   the  universe  at  large.   If  we
extrapolate the  above of hierarchy of  laws to the  large scale, then
the grand unified theory (GUT), when discovered, will be simply be the
tip  of an  indefinite hierarchy  of meta-theories.   We can  then ask
whether a  notion of FW  can, in a  sense, be considered on  a larger,
cosmological scale.   Perhaps spontanenous symmetry  breaking and even
the  Big  Bang,  were  possibly  free  choices by  the  `mind  of  the
universe'.   Finally, given  the greater  computability  supported the
higher-order  laws, it is  impossible for  physical laws  to `compute'
them into existence.  Thus the Big  Bang may well be the `ground zero'
of a Bigger Bang followed by a \textit{Turing cascade}!

\section{Conclusions and discussions \label{sec:conclu}}

We highlighted  the logical paradoxes  encountered in defining  FW and
understanding   its  role   in   physics.   We   highlight  our   main
conclusions. A principal conclusion is  that like the concept of truth
in Tarski's undefinability theorem,  free will is also meta-theoretic:
it correlates freedom to deviate from physical law in a base theory to
parameters of `will' in a meta-theory.  Next, we pointed out that this
meta-theoretic  coupling  must be  continued  infinitely  in order  to
construct the  FW that we intuitively associate  with sentient agents.
Third, the  meta-theories are  shown to correspond  to a  hierarchy of
uncomputability. A  consequence is that  free will is  uncomputable to
infinite degree,  in the  sense that for  a meta-theory at  any finite
order, there  will deviations from  its laws that are  uncomputable in
it.   We justify  this  link between  causality  and computability  by
revisiting self-referential logical paradoxes  that serve as the basis
for theorems of G\"odel-Tarski kind.  Finally, we present a refinement
of  an  anti-mechanist  argument  for  consciousness  based  on  these
theorems.

\end{document}